\documentclass[aps,prd,twocolumn,superscriptaddress,showpacs,showkeys,
nofootinbib]{revtex4-1}

\usepackage{latexsym}
\usepackage{amsmath}
\usepackage{amssymb}
\usepackage{amsfonts}
\usepackage{hyperref}
\usepackage{color}
\usepackage{slashed}
\usepackage{supertabular} 
\usepackage{placeins}
\usepackage{epsfig}
\usepackage{graphicx}
\usepackage{multirow}
\usepackage{epstopdf}

\begin{document}


\title{Symmetries, partners and thresholds: the case of the $X_b$}

\author{Pablo G. Ortega}
\email[]{pgortega@usal.es}
\affiliation{Departamento de Física Fundamental and Instituto Universitario de F\'isica 
Fundamental y Matem\'aticas (IUFFyM), Universidad de Salamanca, E-37008 Salamanca, Spain}

\author{David R. Entem}
\email[]{entem@usal.es}

\author{Francisco Fern\'andez}
\email[]{fdz@usal.es}
\affiliation{Grupo de F\'isica Nuclear and Instituto Universitario de F\'isica 
Fundamental y Matem\'aticas (IUFFyM), Universidad de Salamanca, E-37008 
Salamanca, Spain}

\date{\today}

\begin{abstract}
The discovery of the $X(3872)$ meant the revival of the heavy meson spectroscopy beyond naive $q\bar q$ structures. Once that the $SU(3)$ scheme, which was very useful in the dawn of the quark models, does not work for these states, one has to use  new symmetries, like Heavy Quark Spin Symmetry (HQSS) and Heavy Flavor Symmetry (HFS), to look for new states. However, at the energy regions where these new states appear, new factors are involved  and it is not straightforward to relate  the predictions of the symmetries with the data. In this work, we present a critical analysis of this problem and show, in a coupled-channels model, how the relative position of the bare 
$Q\bar Q$ states with respect to meson-meson thresholds and the coupling with other channels
modulate the strength of the interaction and, hence, modify the structure of the predicted states. We found a possible candidate to the $X(3872)$ partner at $10599$ MeV$/c^2$.
\end{abstract}



\maketitle

\section{Introduction}

History taught us the importance of symmetries in hadron spectroscopy. In fact, the prediction of the existence of the $\Omega^-$ was the first success of the $SU (3)_F$ symmetry and its application allowed to configure a complete hadron map based on the quark model.

The existence of non-conventional quark structures which do not fit in the quark scheme based in the $SU (3)_F$ symmetry is not new, but rolls back to the same origins of the quark model~\cite{GellMann:1964nj}. Although early studies suggested the possible existence of meson-meson molecular states in the charmonium spectrum~\cite{DeRujula:1976zlg},
it was not until 2003, with the observation of the $X(3872)$, when the concept of meson-meson molecule regained 
attention~\cite{Tornqvist:2004qy,Close:2003sg,Voloshin:2003nt}.

The discovery of the $X(3872)$, a $J^{PC}=1^{++}$ state with hidden charm, 
meant the revival of the heavy meson spectroscopy beyond naive $Q\bar Q$ structures. It was first discovered by the Belle Collaboration~\cite{Choi:2003ue} 
and soon confirmed by other Collaborations such as CDF~\cite{Acosta:2003zx}, 
D0~\cite{Abazov:2004kp} and BaBar~\cite{Aubert:2004ns}.
The $X(3872)$ non-conventional properties indicates that more complex structures play 
an active role in the dynamics of the resonance 
(see Ref.~\cite{Guo:2017jvc,Karliner:2017qhf} for more extensive reviews). 

The latest update of its mass throws a value very close to the $D^0\bar D^{\ast\,0}$ threshold, 

\begin{align}
 M_{D^0} + M_{D^{\ast\,0}} - M_{X(3872)} = (0.04 \pm 0.09)\,\text{MeV},
\end{align}
taking PDG average values~\cite{PDG}, which is compatible with a molecular state if the mass is finally confirmed to be below threshold. 

Theoretically, most quark models predict a charmonium state, the $\chi_{c1}(2P)$ state, well above such
threshold, which makes unlikely its assignment to the $X(3872)$. 
Also, some decay properties of the $X(3872)$ are intriguing. The strong decays show a large isospin violation, being
the ratio of isospin-1 decay $J/\Psi\rho^0$, followed by $\rho^0\to\pi^+\pi^-$, similar to the isospin-0 decay to $J/\Psi\omega$, where $\omega\to\pi^+\pi^-\pi^0$,
with value $1.1\pm 0.4$~\cite{PDG}.
This isospin violation is trivially explained with a large $D\bar D^{\ast}$ molecular component in the $X$ wave function and the phase space effect due to the larger width of the $\rho$ meson.

Nevertheless, the charmonium picture is still necessary to explain other observables like the large $\psi(2S) \gamma$ decay with respect to
the $J/\psi \gamma$ decay~\cite{Aaij:2014ala}, suggesting that the wave function of the $X(3872)$ may content a non-negligible charmonium component besides the molecular one.

After the discovery of the $X(3872)$ a great amount of new resonance structures, the so-called XYZ states, have been reported. Then, following the history, one has to look for new symmetries which, based on the measured states, allow to predict new partners that the experimentalists may detect.

Predictions in the charm sector can be connected with the bottom sector assuming that Heavy Flavor Symmetry (HFS) holds. 
This symmetry implies that, in the infinite mass limit, the interaction is the same when you replace the $c$ quark by the $b$ quark. So, those molecules detected in the charm sector are expected to be reproduced in the bottom sector with even larger binding energy,
due to the reduction of the kinetic energy by the larger mass of the $b$ quark. Thus, under this assumption, a partner of the $X(3872)$, called $X_b$, is expected to
lie close to $M_B+M_{B^\ast}\sim 10.6$ GeV, with lower isospin breaking due to the smaller mass splitting between charge and neutral $B^{(\ast)}$ mesons~\cite{Guo:2013sya}.

All those predicted states are based on simple molecular structures and QCD symmetries so, at first, their existence is quite robust.
However, as we have seen in the case of the $X(3872)$, one would expect that nearby $Q\bar Q$ states can mix with these molecular states, which will have an impact on their dynamics, 
changing their composition, binding energy or decay properties~\cite{Entem:2016ojz,Cincioglu:2016fkm}. Moreover, most of the signal reported by the experimentalists lies near meson-meson thresholds and can be interpreted as thresholds cups. In this work, we analyze the influence of all these effects on the predictions of the QCD symmetries. For that purpose, we will study the partners of the $X(3872)$ in $b\bar b$  sectors, using the coupled-channels formalism developed in Ref.~\cite{Ortega:2009hj}, where the $X(3872)$ was found as a $J^{PC}=1^{++}$ $D\bar D^\ast$ state, 
exclusively bound thanks to the coupling with the $2^3P_1$ $c\bar c$, as the direct $D\bar D^\ast$ interaction is not attractive enough to form any bound state.

\section{Symmetries, thresholds and $q\bar q$ states}

The predictions of the different symmetries are not the end of the history. Usually, the XYZ structures are interpreted as threshold cusps. However, it is difficult to identified what and when a threshold is associated with a non-trivial structure. A simple model to answer this question has been suggested in Ref.~\cite{Dong:2020hxe}. Starting with an effective range expansion for the S-wave amplitude in a two-body scattering

\begin{equation}
\label{effective range}
f_0^{-1}(k)=\frac{1}{a_0}+\frac{1}{2} r_0 k^2-ik+\Theta\left(\frac{k^4}{\beta_4}\right)
\end{equation}
where $a_0$ and $r_0$ are the S-wave scattering length and effective range, respectively, $k$ is the center of mass momentum and $\beta$ some hard scale.

One can use the non-relativistic expression for the momentum $k=\sqrt{2\mu E}$ near threshold
and write the amplitude as a function of $E$

\begin{equation}
\label{effective range}
f_0^{-1}(E)=\frac{1}{a_0}-i\sqrt{2\mu E}
\end{equation}
from this expression one gets

\begin{align}
|f_0(E)|^2&=\frac{1}{1/{a_0^2}+2\mu E} & E\geq 0\label{amplitud plus} \\
|f_0(E)|^2&=\frac{1}{\left (1/{a_0}+\sqrt{-2\mu E}\right)^2} & E< 0 \label{amplitud minus}
\end{align}

The half-maximum width of this distribution is

\begin{equation}
\label{width}
\Gamma=\frac{2}{\mu a_0^2}
\end{equation}
which is narrower for large scattering length (strong interaction) and for larger reduced mass (heavy hidden-flavor sector).

For $a_0>0$ (attractive interaction but not enough to form a bound state) the distribution is maximal at $E=0$ and, thus, it appears as a cusp at threshold. Also, there is a virtual pole in the second Riemann sheet of the complex energy plane at $E_{\rm virtual}=-\frac{1}{2\mu a_0^2}$. 

For $a_0<0$, two scenarios are possible. For strong attraction, the pole is located in the first Riemann sheet at $E_{\rm bound}=-\frac{1}{2\mu a_0^2}$ and leads to a near threshold peak. For repulsive interaction, no non-trivial near-threshold peak appears.

Together with the threshold effects, one has to take into account the influence of the $q\bar q$ states near threshold. States with different structures but with the same quantum numbers and similar energies must be coupled. Then, the closest $q\bar q$ states to the two-meson thresholds should be coupled with the two-meson channels~\footnote{In fact, all the $q\bar q$ states with the same quantum numbers should be coupled, but the coupling is negligible for those states which are far from the threshold energy}. Then the hadronic wave  function should be given by

\begin{equation}\label{eq:Wfun}
  | \Psi \rangle = \sum_\alpha c_\alpha |\psi_\alpha\rangle + \sum_\beta \chi_\beta(P)|\phi_{H_Q}\phi_{\bar H_Q}\beta\rangle
\end{equation}
where $|\psi_\alpha\rangle$ are the bare $Q \bar Q$ quark states, $|\phi_{H_Q}\phi_{\bar H_Q}\beta\rangle$ the two-meson states with $\beta$ quantum numbers~\footnote{Naming, for now on, $Q$ as the heavy quarks $c$ and $b$ and
$H_Q(\bar H_Q)$ as the heavy-light meson $D(\bar D)$ or $B(\bar B)$.} and $\chi(P)$ is the relative wave function of the $H_Q \bar H_Q$ channel.

The coupling with the meson spectra induces, in the meson-meson channel, an effective energy-dependent potential given by 

\begin{equation}\label{ec:effectiveV}
  V_{\beta'\beta}^{\rm eff}(P',P) = \sum_\alpha \frac{h_{\beta'\alpha}(P')h_{\alpha\beta}(P)}{E-M_\alpha}
\end{equation}
where $M_\alpha$ are the masses of the bare $Q\bar Q$ mesons and $h_{\alpha\beta}(P)$ is the coupling potential between $Q \bar Q$ and $H_Q\bar H_Q$. 

It is worth noticing that the sign of the effective potential depends whether we are above or below the bare $Q\bar Q$ mass. 
Indeed, for any $H_Q\bar H_Q$ channel coupled to a $Q\bar Q$ state with mass $M_0$, we would have $V_{\rm eff}<0$ (attractive) if $E<M_0$
and $V_{\rm eff}>0$ (repulsive) if $E>M_0$.

Usually, there are several $Q\bar Q$ states below and above threshold and the net attraction or repulsion in the meson-meson channel depends on the balance between the different contributions. Sometimes, it is difficult to generate enough attraction to have a bound state, and threshold cusp linked to virtual states are the most likely explanation for the peaks observed in the experimental data.

This interpretation can explain the results of Ref.~\cite{Entem:2016ojz}, but the situation is even more complicated when several thresholds are involved in the region of interest. In this case, not only the coupling between the bare $Q\bar Q$ and the thresholds has to be considered, but also the non-diagonal elements associated with the coupling channels, which can produce additional attraction. This is the case of the charged resonances $Z_c(3900)^\pm$ or $Z_{cs}(3985)^\pm$, where we do not have $Q\bar Q$ states associated to these resonances but still we get attraction for the coupling between different channels~\cite{Ortega:2018cnm,Ortega:2021enc}.

The conclusion is that symmetries are not enough to predict partners of the well-established states, because the definitions of these states join in other factors like thresholds and meson spectra. In the following, we will show with a well-established quark model how these general considerations work.

 In section II we will describe the model we use, including an analysis of up to what extent the model potential satisfies HQSS. Results for the bottom sector are presented in section III. Finally, we summarized our work in section IV.

\section{The model}

The first ingredient of the present work is the non-relativistic constituent quark model (CQM) extensively described in Ref.~\cite{Vijande:2004he}, which allows us to build the theoretical $Q\bar Q$ and $H_Q$ spectrum and the $H_Q^{(\ast)}\bar H_{Q}^{(\ast)}$ interaction.
The CQM has been successfully employed to explain the hadron phenomenology both in the light and heavy meson 
sectors~\cite{Garcilazo:2001ck,Segovia:2008zza,Segovia:2016xqb} and baryon sectors~\cite{Valcarce:2005em,Ortega:2011zza}, 
from where all the parameters of the model are constrained. Details of the model and explicit expressions 
can be found in Ref.~\cite{Vijande:2004he}, here we will only briefly summarize its most relevant aspects.

The basis of the CQM is the postulation that a constituent mass for quarks emerges as a consequence of the dynamical 
spontaneous chiral symmetry breaking in QCD at some momentum scale. The breaking of the chiral symmetry 
implies the appearance of massless Goldstone bosons ($\pi,K,\eta$). 
The simplest Lagrangian that satisfies the previous properties is,

\begin{equation}
\label{lagrangian}
{\mathcal L}=\overline{\psi }(i\, {\slashed{\partial}} -M(p^{2})U^{\gamma_{5}})\,\psi 
\end{equation}
where $U^{\gamma_5}=e^{i\frac{\lambda _{a}}{f_{\pi }}\phi ^{a}\gamma _{5}}$ is a matrix that codes 
the Goldstone boson fields and $M(p^2)$ is the acquired dynamical constituent mass. 
If this Goldstone boson field matrix $U^{\gamma_{5}}$ is expanded in terms of boson fields, 
we naturally obtain one-boson exchange interactions between quarks. Multi-boson
exchanges are not included, but they are implemented through the exchange of scalar bosons.

The chiral symmetry is explicitly broken in the heavy sector, so Goldstone boson exchanges should not appear among heavy quarks.
However, quarks still interact through a QCD perturbative effect, the gluon exchange diagram~\cite{DeRujula:1975qlm}. Besides, the model incorporates confinement, a non-perturbative QCD effect that avoids colored hadrons. This interaction can be modeled with a screened potential~\cite{Born:1989iv}, which takes into account the
saturation of the potential at some interquark distance due to the spontaneous creation of light-quark pairs (see Refs.~\cite{Valcarce:2005em,Segovia:2013wma} for details).

Meson masses and wave functions are obtained 
by solving the two-body Schr\"odinger equation using the Gaussian Expansion 
Method (GEM)~\cite{Hiyama:2003cu} which is accurate enough and it simplifies 
the subsequent evaluation of the needed matrix elements.
Once we have the internal wave functions of the mesons, we can obtain the interaction between them
using the Resonating Group Method (RGM).
Thus, the $H_Q^{(\ast)}\bar H_{Q}^{(\ast)}$ interaction is given by the so-called RGM direct kernel
\begin{align}\label{ec:RGMkernel}
    ^{RGM}V_D(\vec{P}',\vec{P}) = 
    \sum_{i,j} \int d\vec{p}_{\xi_C}d\vec{p}_{\xi_D}d\vec{p}_{\xi_A} d\vec{p}_{\xi_B} \nonumber\\
    \phi ^*_C (\vec{p}_{\xi_C}) \phi ^*_D (\vec{p}_{\xi_D}) 
    V_{ij}(\vec{P}',\vec{P}) \phi _A (\vec{p}_{\xi _A}) \phi _B (\vec{p}_{\xi _B}).
\end{align}
for a general $AB\to CD$ process.

Actually, quarks or antiquarks exchanges between different mesons are allowed. 
Such interactions couple different meson states, such as, for instance, the $D\bar D^*\to J/\Psi\omega$ channels. 
However, this sort of processes are suppressed by the meson wave functions overlaps.

As mentioned in the introduction, nearby $Q\bar Q$ states to the two meson thresholds can have an important effect
in the dynamics of the system. As we will see later, the closer the state, the larger the effect. Hence, the second ingredient of our study is
a mechanism that can couple two- and four-quark states.
The coupling between the two sectors require the creation of a light $q\bar q$ pair. In principle,
this process can be deduced from the same quark-quark interaction that drives the meson dynamics. 
However, the quark pair creation $^3P_0$ model~\cite{Micu:1968mk,LeYaouanc:1972vsx} provides similar results to those microscopic calculations
within a simpler approach, as shown by Ref.~\cite{Ackleh:1996yt}. The non-relativistic reduction of the $^3P_0$
Hamiltonian is equivalent to the transition operator~\cite{Bonnaz:1999zj},

\begin{align}
\mathcal{T}=&-24 \pi^{1/2}\gamma\sum_\mu \int d^3 p d^3p' \,\delta^{(3)}(p+p')\nonumber \\
&\left[ \mathcal Y_1\left(\frac{p-p'}{2}\right) b_\mu^\dagger(p)
d_\nu^\dagger(p') \right]^{C=1,I=0,S=1,J=0}
\label{TBon}
\end{align}
being $\mu$ ($\nu=\bar \mu$) the $q$ ($\bar q$) quantum numbers and
$\gamma= g/(2m)$ a dimensionless parameter that controls the $q\bar q$ pair creation strength from the vacuum.
The transition potential $h_{\beta \alpha}(P)$ within the $^3 P_0$ model can be expressed as
\begin{equation}
\label{Vab}
        \langle \phi_{H_Q} \phi_{\bar H_Q} \beta | \mathcal{T}| \psi_\alpha \rangle =
        P \, h_{\beta \alpha}(P) \,\delta^{(3)}(\vec P_{\mbox{cm}})
\end{equation}
where $P$ is the relative momentum of the two meson state $H_Q \bar H_Q$,
$|\psi_\alpha\rangle$ are the $Q\bar Q$ hidden heavy mesons and $|\phi_{H_Q}\phi_{\bar H_Q}\beta\rangle$ are the two
meson states with $\beta$ quantum numbers.

It is worth noticing that the model is controlled by only one parameter, the coupling $\gamma$. 
The value of the $\gamma$ for the charmonium sector was constrained in Ref.~\cite{Ortega:2012rs} for the $X(3872)$.
However, such value does not necessarily have to be the same for other sectors. Indeed, an overall
good description of the two meson strong decays for different sectors is obtained if the $\gamma$ is 
logarithmically scaled with the reduced mass of the two quarks of the decaying meson, as analyzed in Ref.~\cite{Segovia:2012cd},
A satisfactory agreement was obtain, as well as constrains on the value of the $\gamma$, which will be employed in this work for the $b\bar b$ sector.
Additionally, in order to quantify the sensitivity of the results with the value of $\gamma$, 
a variation of $10\%$ will be included in this parameters. 
Thus, the values of the $^3P_0$ parameter $\gamma$ used for the bottomonium sector is $\gamma= 0.205\pm 0.020$ .

In order to perform a full coupled-channels calculation between the $Q\bar Q$ and the $H_Q^{(\ast)}\bar H_Q^{(\ast)}$  channels
we follow Ref.~\cite{Ortega:2012rs} (all the details can be found therein). We first assume the combination for the wave function given in Eq.~\eqref{eq:Wfun}.
We recall that all the two-body wave functions are obtained from the solution of the two-body problem with CQM quark-quark 
interactions, expressed with the GEM.

Gathering the RGM direct kernels obtained from RGM (Eq.~\ref{ec:RGMkernel}) and the coupling with $Q\bar Q$ bare mesons, we obtain a 
coupled-channels equation for the relative wave function of the two mesons:
\begin{align}\label{ec:Ec1}
\sum_{\beta}\int H_{\beta'\beta}(P',P) \chi_{\beta}(P) {P}^2 dP = E \chi_{\beta'}(P'),
\end{align}
with $H_{\beta'\beta}(P',P)=H^{\rm RGM}_{\beta'\beta}(P',P)+V^{\rm eff}_{\beta'\beta}(P',P)$.
In the latter equation, we have denoted $H^{\rm RGM}_{\beta'\beta}$ as the RGM Hamiltonian for the two meson
states obtained from the underlying $q-q$ interaction and $V^{\rm eff}_{\beta'\beta}(P',P)$ is the effective energy-dependent potential between the two mesons, expressed as
Eq.~\eqref{ec:effectiveV}, induced by the coupling with $Q\bar Q$ states.

This approach can describe both the renormalization of the bare $Q\bar Q$ states due to the presence of nearby
meson-meson thresholds and the generation of new states through the meson-meson interaction due to the coupling with $Q\bar Q$
states and the underlying $q$-$q$ interaction, as it is the case for the $X(3872)$ in our model~\cite{Ortega:2009hj}.

\begin{figure}[!t]
\centering
\includegraphics[width=0.5\textwidth]{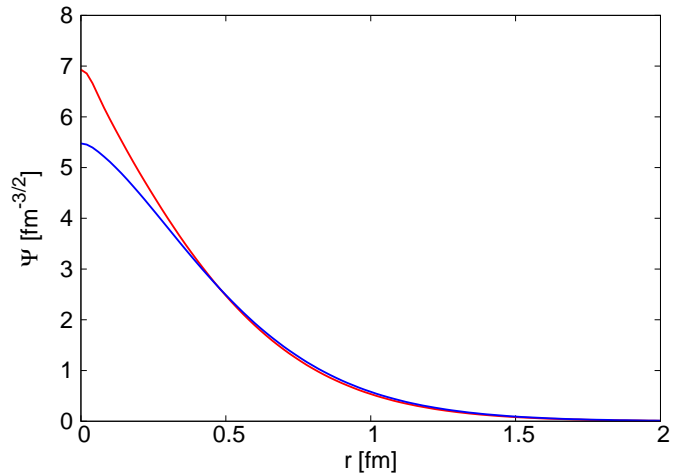}
\caption{\label{fig:HQWF} Radial wave function in coordinate space for $B$ (red line) and $B^\ast$ (blue line). The HQSS breaking due to the spin-spin term arises in the $r<0.5$ fm region.}
\end{figure}

Before presenting the results, it is worth exploring the level of agreement of our RGM kernels to HQSS.
Indeed, some breaking is expected due to the 
finite heavy quark masses, of order $\mathcal{O}(m_Q^{-1})$, although they will introduce a small effect as the heavy quark masses are large.
This HQSS breaking is shown in Figs.~\ref{fig:HQWF} for the radial wave functions of the $B$ and $B^\ast$ mesons in coordinate space.
Within our model, the potential that regulates such breaking is the spin-spin term of the one-gluon-exchange interaction,

\begin{equation}
 \Delta V_{\rm OGE}^{D^\ast-D}(\vec r_{ij})=-\frac{1}{24m_Q^2}\alpha_s\left(\lambda_i^c\cdot\lambda_j^c\right)\frac{e^{-r_{ij}/r_0(\mu)}}{r_{ij}r_0^2(\mu)}
\end{equation}
which goes as $\mathcal{O}(m_Q^{-2})$.

For exact HQSS, these wave functions should be the same and the $S$-wave two-meson state potentials should satisfy the following relations,

\begin{align}
V_{H_Q\bar H_Q^\ast\to H_Q\bar H_Q^\ast}^{1^{++}} &= V_{H_{Q}^\ast\bar  H_{Q}^\ast\to H_{Q}^\ast\bar  H_{Q}^\ast}^{2^{++}}=\nonumber \\
 &= \frac{3}{2}V_{H_Q\bar H_Q\to H_Q\bar H_Q}^{0^{++}}-\frac{1}{2}V_{H_Q^\ast\bar  H_Q^\ast\to H_Q^\ast\bar  H_Q^\ast}^{0^{++}}.
\label{Ec5}
\end{align}

\begin{figure}[!t]
\centering
\includegraphics[width=0.5\textwidth]{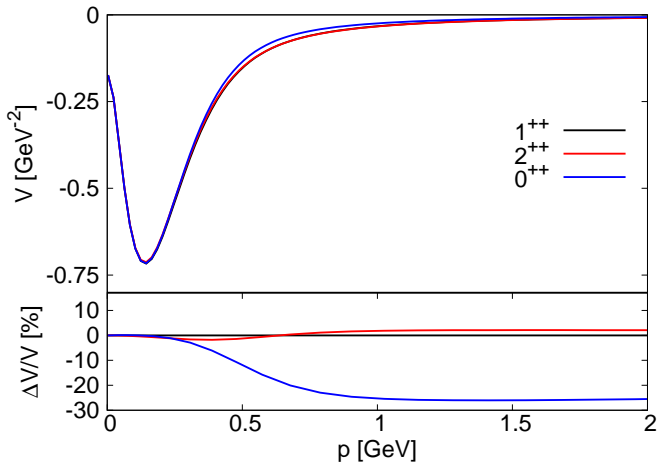}
\caption{\label{fig:Pot1} Examples of HQSS fulfillment for some RGM kernels considered in this work (Eq.~\eqref{Ec5}: $V_{B\bar B^\ast\to B\bar B^\ast}^{1^{++}}(p,p)$ (black line), $V_{B^\ast \bar B^\ast\to B^\ast\bar  B^\ast}^{2^{++}}(p,p)$ (red line) and $\frac{3}{2}V_{B\bar B\to B\bar B}^{0^{++}}(p,p)-\frac{1}{2}V_{B^\ast\bar  B^\ast\to B^\ast\bar  B^\ast}^{0^{++}}(p,p)$ (blue line).
The lower pannel shows the difference (in \%) of the previous kernels with respect to $V_{B\bar B^\ast\to B\bar B^\ast}^{1^{++}}(p,p)$. }
\end{figure}

However, as one can see in Fig.~\ref{fig:Pot1}, a small HQSS breaking effect is induced by the small difference 
in the wave functions of the pseudoscalar and vector heavy mesons. Such breaking is below $2\%$ for the $1^{++}$ vs $2^{++}$ sector, but larger for the $0^{++}$ sector at large momentum.
One should expect this behaviour, since HQSS breaking terms are short-range $m_Q^{-1}$ terms, and so more important
for large $p$.
Between these limits, the HQSS symmetry suggests similar results for these three sectors.

\section {Results}

As shown in Fig.~\ref{fig:Pot1}, our model satisfies HQSS, despite the slight breaking due to the finite value of the heavy quark mass. 
However, the relation among bare $Q\bar Q$ states and meson-meson thresholds vary for different $J^{PC}$ quantum numbers.
Thus, extrapolations on the existence of spin or heavy partners of the well-established state  based solely
on HQSS assumptions should be taken with caution.

In this work we will focus on the bottomonium sector, as the charmonium one has been widely studied in Refs.~\cite{Ortega:2009hj,Ortega:2012rs,Ortega:2017qmg}. We consider all the $b\bar b$ states, predicted by CQM, within $\pm 150$ MeV around the closest open $B^{(\ast)}\bar B^{(\ast)}$ threshold in $S$ or $D$ wave (Table~\ref{tab:thresholds} shows the mass of the considered thresholds and Table~\ref{tab:CQMbare} the theoretical $b\bar b$ states.).
The effect of farthest thresholds in the $b\bar b$ spectra is smooth and we expect it to be encoded in the screened confinement potential as a global contribution. 
Hence, we will consider the channels:

\begin{enumerate}
\item $J^{PC}=0^{++}$: $B \bar B$ ($^1S_0$), $B^*\bar B^*$ ($^1S_0$-$^5D_0$) and $\omega\Upsilon(2S)$ ($^1S_0$-$^5D_0$).
\item $J^{PC}=1^{++}$: $B \bar B^*$ ($^3S_1$-$^3D_1$), $B^*\bar B^*$ ($^5D_1$) and $\omega\Upsilon(2S)$ ($^3S_1$-$^3D_1$-$^5D_1$).
\item $J^{PC}=2^{++}$: $B\bar B$($^1D_2$), $B\bar B^*$ ($^3D_2$), $B^*\bar B^*$ ($^5S_2$-$^1D_2$-$^5D_2$) and $\omega\Upsilon(2S)$ ($^5S_2$-$^1D_2$-$^3D_2$-$^5D_2$).
\end{enumerate}
where the partial waves are in parenthesis. The $\omega\Upsilon(1S)$ channel is too far below to have a significant contribution to the coupled-channel calculation, but we will calculate perturbatively the decay of the resulting states to the latter one. As the charge to neutral mass of the $B^{(\ast)}$ mesons is small, isospin-breaking effects are expected to be negligible, so they will not be included.

\begin{table}
 \caption{\label{tab:thresholds} Energies [MeV] of the isospin-averaged $B^{(\ast)}\bar B^{(\ast)}$ thresholds, from PDG~\cite{PDG}.}
 \begin{tabular}{c|ccc}
  \hline
  \hline
  Channel & $B\bar B$ &  $B\bar B^\ast$ & $B^\ast\bar B^\ast$\\  \hline
  Energy  & 10558.49&   10604.44     &  10650.20  \\
  \hline
  \hline
 \end{tabular}
\end{table}

\begin{table}
 \caption{\label{tab:CQMbare} Theoretical bare $b\bar b$ masses (in MeV) within $\pm 150$ MeV from 
 $B^{(\ast)}\bar B^{(\ast)}$ thesholds of Tab.~\ref{tab:thresholds}, obtained from CQM predictions~\cite{Segovia:2008zz,Segovia:2016xqb}.}
 \begin{tabular}{ccrr}
  \hline
  \hline
  $J^{PC}$  & $n^{2S+1}L_J$ & Mass  \\
  \hline
 $0^{++}$  & $3^3P_0$ & 10499.96\\
           & $4^3P_0$ & 10726.18\\
 $1^{++}$  & $3^3P_1$ & 10512.76 \\
           & $4^3P_1$ & 10737.27 \\
 $2^{++}$  & $3^3P_2$ & 10520.89 \\
           & $2^3F_2$ & 10569.48 \\
           & $4^3P_2$ & 10744.37 \\
           & $3^3F_2$ & 10781.08 \\                                                          
  \hline
  \hline
 \end{tabular}
\end{table}

\begin{table}
\caption{\label{tab:b0} Dressed $b\bar b$ states  for $J^{PC}=0^{++}$}
\begin{tabular}{l|cc|cc}
\hline\hline
     & \multicolumn{2}{|c|}{State 1} & \multicolumn{2}{|c}{State 2} \\
\hline
Mass [MeV] & \multicolumn{2}{|c|}{$10458.0$} & \multicolumn{2}{|c}{$10753.4$}\\
Width [MeV] & \multicolumn{2}{|c|}{$0.001$} & \multicolumn{2}{|c}{$36.3$} \\\cline{2-5}
& ${\cal P} [\%]$ & $\Gamma$ [MeV] & ${\cal P} [\%]$ & $\Gamma$ [MeV]\\\cline{2-5}
$b\bar b(3^3P_0)$ & $87.83$ & $-$ & $4.52$ & $-$\\
$b\bar b(4^3P_0)$ & $0.03$  & $-$ & $24.92$ & $-$\\
$B\bar B$ & $4.56$& $0$ & $17.03$ & $2.93$\\
$B^*\bar B^*$ & $7.58$& $0$ & $53.22$ & $31.13$ \\
$\Upsilon(2S)\omega$ & $0.00$ & $0$ & $0.31$ &$ 0$\\ \hline
$\Upsilon(1S)\omega$ & $-$ & $0.001$ & $-$ & $2.25$ \\
\hline\hline
 \end{tabular}
\end {table}

\begin{table*}
\caption{\label{tab:b1} Dressed $b\bar b$ and additional states for $J^{PC}=1^{++}$}
\begin{tabular}{l|cc|cc|cc|cc}
\hline\hline
     & \multicolumn{2}{|c|}{State 1} & \multicolumn{2}{|c|}{State 2} & \multicolumn{2}{|c|}{State 3} & \multicolumn{2}{|c}{State 4}\\
\hline
Mass [MeV] & \multicolumn{2}{|c|}{$10471.9$} & \multicolumn{2}{|c|}{$10599.3$}   & \multicolumn{2}{|c|}{$10738.8$}  & \multicolumn{2}{|c}{$10759.5$}\\
Width [MeV] & \multicolumn{2}{|c|}{$0$} & \multicolumn{2}{|c|}{$0.51$}  & \multicolumn{2}{|c|}{$0.56$}& \multicolumn{2}{|c}{$13.51$}\\\cline{2-9}
& ${\cal P} [\%]$ & $\Gamma$ [MeV] & ${\cal P} [\%]$ & $\Gamma$ [MeV]& ${\cal P} [\%]$ & $\Gamma$ [MeV]& ${\cal P} [\%]$ & $\Gamma$ [MeV] \\\cline{2-9}
$b\bar b(3^3P_1)$ & $88.70$ & $-$ & $0.33$ & $-$  & $0.08$ & $-$& $2.97$ & $-$\\
$b\bar b(4^3P_1)$ & $0.02$  & $-$ & $0.14$ & $-$ & $0.96$ & $-$& $11.82$ & $-$\\
$B\bar B^*$ & $6.26$ & $0$ & $93.65$ & $0$  & $20.01$ & $0.47$& $48.09$ & $4.30$\\
$B^*\bar B^*$ & $5.02$& $0$ & $5.84$ & $0$  & $78.94$ & $0.07$& $37.02$ & $4.30$\\
$\Upsilon(2S)\omega$ & $0.00$ & $0$ & $0.04$ &$0$ & $0.003$ & $0$& $0.002$ & $0$\\ 
$\Upsilon(1S)\omega$ & $-$ &$5\cdot 10^{-4}$ & $-$ & $0.51$  & $-$ & $0.015$& $-$ & $4.91$\\\hline
\hline\hline
 \end{tabular}
\end{table*}

\begin{table*}
\caption{\label{tab:b2} Dressed  $b\bar b$ states and additional for $J^{PC}=2^{++}$}
\begin{tabular}{l|cc|cc|cc|cc|cc}
\hline\hline
     & \multicolumn{2}{|c|}{State 1} & \multicolumn{2}{|c|}{State 2} & \multicolumn{2}{|c|}{State 3}& \multicolumn{2}{|c|}{State 4}& \multicolumn{2}{|c}{State 5} \\
\hline
Mass [MeV] & \multicolumn{2}{|c|}{$10485.2$}& \multicolumn{2}{|c|}{$10551.6$} & \multicolumn{2}{|c|}{$10685.6$}& \multicolumn{2}{|c|}{$10766.9$}& \multicolumn{2}{|c}{$10768.3$}\\
Width [MeV] & \multicolumn{2}{|c|}{$0$} & \multicolumn{2}{|c|}{$0$}& \multicolumn{2}{|c|}{$32.19$}& \multicolumn{2}{|c|}{$13.3$}& \multicolumn{2}{|c}{$65.48$} \\\cline{2-11}
& ${\cal P} [\%]$ & $\Gamma$ [MeV] & ${\cal P} [\%]$ & $\Gamma$ [MeV]& ${\cal P} [\%]$ & $\Gamma$ [MeV]& ${\cal P} [\%]$ & $\Gamma$ [MeV]& ${\cal P} [\%]$ & $\Gamma$ [MeV]\\\cline{2-11}
$b\bar b(3^3P_2)$ & $88.56$ & $-$& $1.14$ & $-$  & $0.26$ & $-$ & $0.46$ & $-$ & $2.39$ & $-$\\
$b\bar b(2^3F_2)$ & $0.50$ & $-$& $83.59$  & $-$ & $3.02$ & $-$  & $0.13$ & $-$ & $0.07$ & $-$ \\
$b\bar b(4^3P_2)$ & $0.02$ & $-$& $0.01$ & $-$ & $23.00$ & $-$ & $3.66$ & $-$ & $15.93$ & $-$\\
$b\bar b(3^3F_2)$ & $0.00$ & $-$& $0.05$  & $-$ & $6.05$ & $-$& $33.91$ & $-$  & $17.79$ & $-$\\
$B\bar B$ & $2.87$ & $0$& $10.92$ & $0$ & $13.47$ & $4.49$& $21.21$ & $4.52$  & $17.32$ & $37.3$\\
$B\bar B^*$  & $1.83$ & $0$& $3.08$ & $0$& $4.57$ & $14.97$& $12.93$ & $4.05$  & $2.42$ & $2.88$\\
$B^*\bar B^*$  & $6.23$ & $0$& $1.22$& $0$ & $49.57$ & $12.62$ & $27.68$ & $4.12$ & $44.05$ & $25.25$\\
$\Upsilon(2S)\omega$ & $0.00$ & $0$ & $0.00$ &$ 0$& $0.05$ & $0.00$& $0.01$ & $0$  & $0.03$ & $0.00$\\ \hline
$\Upsilon(1S)\omega$ & $-$ &$0$ & $-$ & $0$& $-$ & $0.11$& $-$ & $0.61$  & $-$ & $0.05$\\
\hline\hline
 \end{tabular}
\end{table*}

The masses of the thresholds energies and the bare $q\bar q$ states are listed in Table~\ref{tab:thresholds} and Table~\ref{tab:CQMbare}.
The results for dressed $b\bar b$ states and additional ones are shown in Table~\ref{tab:b0}, Table~\ref{tab:b1} and Table~\ref{tab:b2} for $J^{PC}=0^{++}$, $1^{++}$ and $2^{++}$, respectively.

\begin{figure}[!t]
\centering
\includegraphics[width=0.5\textwidth]{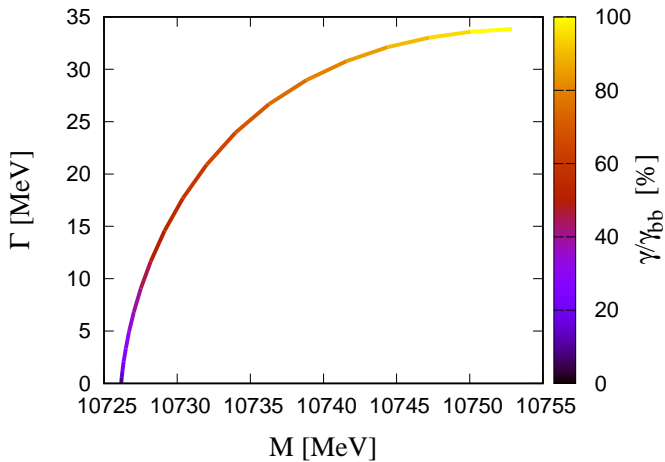}
\caption{\label{fig:0pp} Trajectories of the $J^{PC}=0^{++}$ state above the $B^\ast\bar B^\ast$ threshold with increasing values of the $^3P_0$ strength constant $\gamma$. Results are given in \% with respect to the value $\gamma_{bb}=0.205$ of the $b\bar b$ sector.}
\end{figure}

As a result of the calculation in  the $0^{++}$ sector, we obtain two states. The first one is basically a $3^3P_0$ $q\bar q$ state ($87\%$)
renormalized by the coupling with the nearest thresholds.  The second one is compatible with a $4^3P_0$ state, but with an important $B^*\bar B^*$ component. In Fig.~\ref{fig:0pp}, the evolution of this state with increasing values of $\gamma$ is shown. This is an example that states which are above threshold, like the $4^3P_0$, are more than simple $q\bar q$ states. However, it does not appear any extra molecular states besides the $3^3P_0$ and the $4^3P_0$. Our naive analysis of the relative position of the $q\bar q$ states with respect to the threshold would suggest a $B\bar B$ molecular state, because the $4^3P_0$ adds an extra attraction to the $q$-$q$ interaction, whereas the $3^3P_0$  contribute with much less repulsion. But, in this particular channel, the one-pion exchange interaction is forbidden for the $B\bar B$ and the $q$-$q$ interaction is not attractive enough to form any extra states.

In the $1^{++}$ sector we obtain four states (Table~\ref{tab:b1}). The first one corresponds with a renormalized  $3^3P_1$ state ($88.7\%$) at $10471.9$ MeV/$c^2$, compatible with the experimental mass of the $\chi_{b1}(3P)$  ( $10512.1\pm 2.3$ ) within the uncertainties of the model. A second state, at $10759.54$ MeV/$c^2$, with a sizable $4^3P_1$ component ($12\%$) would correspond to the $\chi_{b1}(4P)$, although, once again, it is shown that states above threshold have a very complex structure.

\begin{figure}[!t]
\centering
\includegraphics[width=0.5\textwidth]{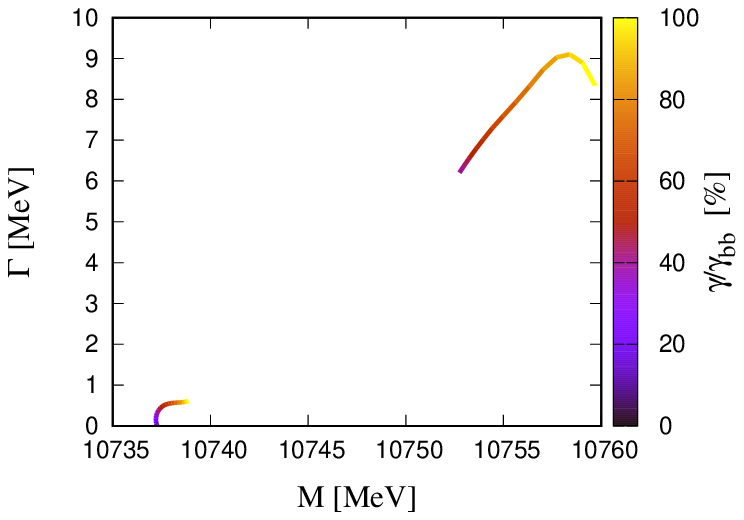}
\caption{\label{fig:1pp} Trajectories of the $J^{PC}=1^{++}$ states above the $B\bar B^\ast$ threshold with increasing values of the $^3P_0$ strength constant $\gamma$. Same legend as in Fig.~\ref{fig:0pp}.}
\end{figure}

Besides, two new states emerge. The first one is basically a $B\bar B^*$ molecule ($93.65\%$) at an energy of $10599$ MeV/$c^2$, whereas the second correspond to a $B^*\bar B^*$ molecule at $10738$ MeV/$c^2$. Although this state is forbidden in S-wave, its 
$^5D_1$ can coupled to the $B\bar B^*$ channel through the tensor interaction of the pion, which produces enough attraction to obtain a resonant state. In Fig.~\ref{fig:1pp}, we show the trajectory of the two states above the $B^*\bar B^*$ threshold with increasing values of $\gamma$.

\begin{figure}[!t]
\centering
\includegraphics[width=0.5\textwidth]{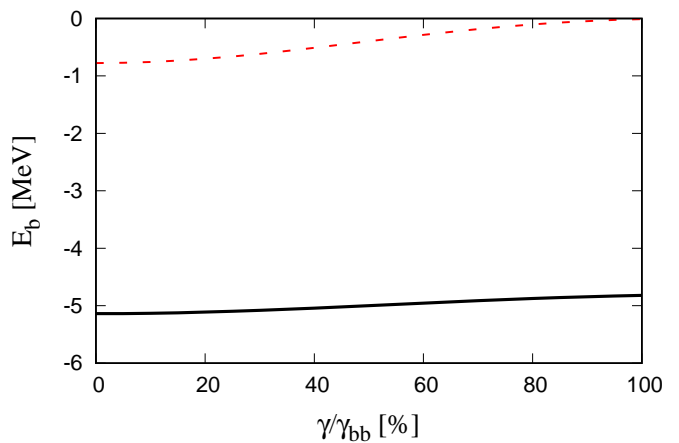}
\caption{\label{fig:Bound1pp} Binding energy of the bottom partner of the $X(3872)$ as a function of the $^3P_0$ strength constant $\gamma$, expressed in \% with respect to the $b\bar b$ sector value of $\gamma_{bb}=0.205$. The black-solid line shows the full coupled-channels calculation including $B\bar B^*+B^*\bar B^*+\Upsilon(2S)\omega$, whereas the red-dashed shows the binding energy only including the $B\bar B^*$ channel.}
\end{figure}

Lets study with more detail the $10599$ MeV/$c^2$ state. In our naive approach, it is a candidate to a bound state or resonance, because
the $4^3P_1$ state is $133$ MeV/$c^2$ above threshold whereas the $3^3P_1$ state is $92$ MeV/$c^2$  below threshold. Contrary to the $X(3872)$ case, this configuration  will give repulsion according with Eq.~\eqref{ec:effectiveV}. However, as we can see in Fig.~\ref{fig:Bound1pp}, the repulsion is not enough to unbound the $B\bar B^*$ molecule.
In fact, it is the coupling with the $B^*\bar B^*$ channel which mostly brings the state below threshold. 
It is worth reminding that the $B^*\bar B^*$ threshold is around $45$ MeV above the $B\bar B^*$ one, whereas in the charm sector $m_{D^*D^*}-m_{DD^*}\approx 140$ MeV, so the influence of the $H_Q^*\bar H_Q^*$ channel in the bottom sector is larger than in the charm one. The prediction is robust because it does not depend of the values of $\gamma$. 
In Table~\ref{tab:Xb} we show the properties of the $X_b$ considering a $10\%$ uncertainty in the value of the $\gamma$ parameter. That is an example that the final result is not a simple consequence of the symmetry or the meson spectrum, but also, the nearby coupled channels play an important role to define the energy of the state or the resonance. The mass of this state agrees with the estimations of Ref.~\cite{Guo:2013sya} and~\cite{Karliner:2013dqa}.

Searches for the $X_b$ partner, carried out by the CMS and ATLAS Collaboration in the $\Upsilon\pi^+\pi^-$ channel~\cite{Chatrchyan:2013mea,ATLAS:2014mka}, analog to the $J/\Psi\pi^+\pi^-$ decay of the $X(3872)$, and in the $J/\Psi\pi^+\pi^-\pi^0$ channel by Belle have been unfruitful to date. 
Nevertheless, this setback does not rule out its existence, as the expected lower isospin breaking due to the smaller mass splitting between charged and neutral $B^{(*)}$ mesons, contrary to the $X(3872)$ case, leads to a strong suppression to the $J/\Psi\pi^+\pi^-$ channel. Concerning the $J/\Psi\pi^+\pi^-\pi^0$ channel, assuming that the three pions comes from the $\omega$ meson, one can see from Table~\ref{tab:Xb} that the $\Upsilon \omega$ channel is weakly coupled to the $X_b$. However, it still remains as the best channel to detect the elusive $X_b$, because
the radiative decays $\Upsilon(1S)\gamma$ and $\Upsilon(2S)\gamma$, calculated through the standard expressions for the electric dipole transition~\cite{Brambilla:2010cs} of the $b\bar b$ components of the $X_b$, are negligible (see Table~\ref{tab:Xb}). Therefore, further experimental searches should explore other channels such as the
$B\bar B^*$, which could be achieved in the future by the SuperKEKB project.

\begin{figure}[!t]
\centering
\includegraphics[width=0.5\textwidth]{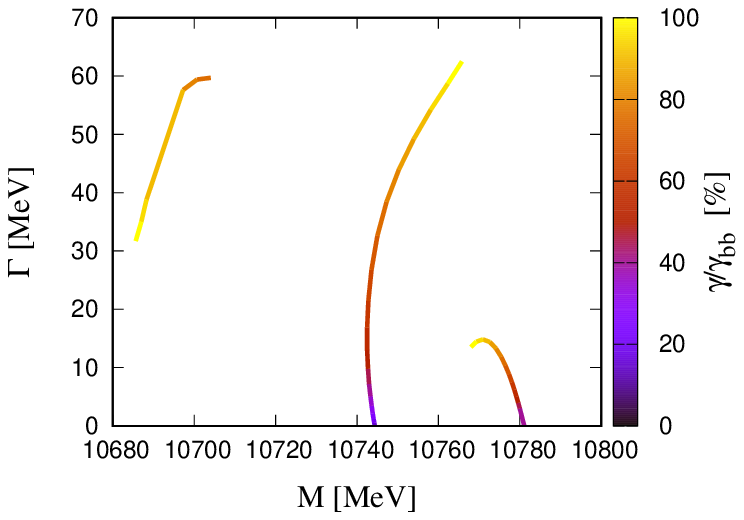}
\caption{\label{fig:2pp} Trajectories of the $J^{PC}=2^{++}$ states above the $B\bar B^\ast$ threshold with increasing values of the $^3P_0$ strength constant $\gamma$. Same legend as in Fig.~\ref{fig:0pp}.}
\end{figure}

Finally, in the $2^{++}$ sector, we found up to five states. Two of them are below the $B\bar B$ threshold, which correspond to the renormalized $2^3F_2$ and $3^3P_2$ $q\bar q$ states. The third and fourth ones, at $10685$ MeV/$c^2$ and $10766.9$ MeV/$c^2$, can be identified with a $4^3P_2$ and $3^3F_2$ $q\bar q$ states, respectively, with an important component of $B^*\bar B^*$ molecule. The last one, at $10768.3$ MeV/$c^2$, has an important $B^*\bar B^*$ component with sizable contributions of the $4^3P_2$ and $3^3F_2$ $q\bar q$ states. In Fig.~\ref{fig:2pp}, we show the trajectory of the states above the $B^*\bar B^*$ threshold with increasing values of the $^3P_0$ parameter $\gamma$. As in the other sectors, states which are below threshold can be clearly identified with renormalized $q\bar q$ states while states above threshold acquire important molecular component which makes it hard to identify them with pure $q\bar q$ states.

\begin{table*}
\caption{\label{tab:Xb} Stability of the properties of the $J^{PC}=1^{++}$ $X_b$ candidate for $\gamma=0.205\pm0.020$. The probabilities are given in \% and the partial widths in MeV.}
\begin{tabular}{cc|ccccc|ccc}
\hline\hline
Mass [MeV] & $\Gamma$ [MeV] & ${\cal P}_{b\bar b(3^3P_1)}$ & ${\cal P}_{b\bar b(4^3P_1)}$ & ${\cal P}_{B\bar B^*}$ & ${\cal P}_{B^*\bar B^*}$ & ${\cal P}_{\Upsilon(2S)\omega}$ & $\Gamma_{\Upsilon(1S)\omega}$ & $\Gamma_{\Upsilon(1S)\gamma}$ & $\Gamma_{\Upsilon(2S)\gamma}$   \\ \hline
$10599.30_{-0.02}^{+0.01}$  & $0.51\pm 0.01$ & $0.33\pm0.02$ & $0.14_{-0.03}^{+0.05}$ & $93.65_{-0.07}^{+0.08}$ & $5.84_{-0.04}^{+0.01}$ & $0.04\pm 0.01$ & $0.51\pm 0.01$ & $(2.5\pm0.2)\cdot 10^{-5}$ &  $(2.8\pm 0.2)\cdot 10^{-5}$  \\
\hline\hline
 \end{tabular}
\end{table*}

\section{Conclusions}

Symmetries have played an important role in the development of the hadron spectroscopy. That is the reason why when a new and unexpected state appears, like the $X(3872)$, one is tempted to use symmetries (SU(3), HQSS or HFS) to predict new resonances. In this work we show that, regardless of the final result, this extrapolation is not straightforward, because when we are close to the meson-meson thresholds new dynamics appear which can modified the symmetries predictions. We perform, in the bottom sector, coupled-channels calculations in which both $b\bar b$ states and meson-meson channels are taken into account. Although the original model satisfy HQSS symmetry, the coupling with $b\bar b$ states modifies the $q$-$q$ potential depending on the relative position of these states with respect to the thresholds. Furthermore, the nondiagonal elements between different meson-meson channels would also modify the interaction. As a sum of all these effects, one can conclude that the nature of the resulting states is more complicated than the estimations based on HFS/HQSS symmetries.

We have analyzed the $J^{PC}=0^{++},1^{++}$ and $2^{++}$ $B^{(*)}\bar B^{(*)}$ coupled with all the $b\bar b$ states, predicted by CQM, within $\pm 150$ MeV around the closest open $B^{(\ast)}B^{(\ast)}$ threshold in $S$ or $D$ wave. Most of the states can be indentified with $b\bar b$ states renormalized by the coupling with the meson-meson channels. This renormalization is more important for states above threshold, where the coupling with the molecular components represent more than $50\%$ of the composition of the states.

Only two new states appear, with quantum numbers $J^{PC}=1^{++}$. The first one is basically a $B\bar B^*$ molecule ($93.65\%$) at an energy of $10599$ MeV/$c^2$, whereas the second correspond to a $B^*\bar B^*$ molecule at $10738$ MeV/$c^2$. Further experimental searches, possibly looking to the $B\bar B^*$ channel, would confirm the existence of the $X_b$ molecule.


\begin{acknowledgments}
This work has been partially funded by 
EU Horizon2020 research and innovation program, STRONG-2020 project, under grant agreement no. 824093 and
Ministerio Espa\~nol de Ciencia e Innovaci\'on, grant no. PID2019-105439GB-C22/AEI/10.13039/501100011033.
\end{acknowledgments}


\bibliography{Symmetries}

\end{document}